# Dipole Anisotropy from an Entropy Gradient

David Langlois[1,2] and Tsvi Piran[1]

[1] *Racah Institute of Physics, The Hebrew University,*

*Givat Ram, 91904 Jerusalem, Israel.*

[2] *Département d'Astrophysique Relativiste et de Cosmologie, C.N.R.S.,*

*Observatoire de Paris, section de Meudon, 92195 Meudon, France.*

## Abstract

It is generally accepted that the observed CMBR dipole arises from the motion of the local group relative to the CMBR frame. An alternative interpretation is that the dipole results from an ultra-large scale ($\lambda > 100 c/H_0$) isocurvature perturbation. Recently it was argued that this alternative possibility is ruled out. We examine the growth of perturbations on scales larger than the Hubble radius and in view of this analysis, we show that the isocurvature interpretation is still a viable explanation. If the dipole is due to peculiar motion then it should appear in observations of other background sources provided that they are distant enough.

## I. INTRODUCTION

The dipole moment is the most prominent feature in the Cosmic Microwave Background Radiation (CMBR) anisotropy [1]. The dipole, which is larger by two orders of magnitude than all other multipoles, is generally accepted to result from the earth's motion relative to the CMBR frame. The main purpose of this paper is to emphasize the fact that the origin of the dipole is not *necessarily* a Doppler effect. This idea has already been put forward by a few authors ( [2], [3], [4]). However, it was recently claimed that these arguments were wrong [5]. We show here that a large scale isocurvature model for the dipole is a viable alternative to the Doppler origin. Current observations of the CMBR dipole and quadrupole are consistent with this possibility. We examine other potential implications of this scenario and suggest observations that would confirm or rule out the Doppler origin of the dipole.

Our starting point here will be the paper by Paczynski and Piran [3], hereafter denoted PP. This paper is based on a Tolman-Bondi model (spherical symmetry and dust), which contains a (gravitationally negligible) spherical distribution of radiation. It is shown that a non centered observer can measure a significant dipole due to such a radially varying specific entropy (i.e. the ratio of the number density of photons to the number density of baryons). These results were obtained by integrating numerically the light geodesics in the Tolman-Bondi model.

The phenomenon described in PP can appear to the reader a bit artificial by the choice of a very particular space-time model and the results are not intuitive in view of the complicated numerical integration involved. Moreover, there was a recent claim [5] that the results of PP are wrong and that it is impossible to obtain a dipole far larger than the quadrupole from either isocurvature or adiabatic perturbations. It is, therefore, the purpose of this paper to demonstrate that the phenomenon described in PP is on one hand true and on the other hand is more general than it seems at first glance. To do so we show that the results of PP can be obtained in the context of a linearly perturbed Friedmann-Robertson-Walker (FRW) universe, within which the computation of the dipole and of the quadrupole can be carried out analytically. These results depend, in fact, only on one crucial argument: the presence of very large scale isocurvature perturbations (by very large scales, we mean scales far larger than the Hubble radius today). It is essential to stress that the modes that contribute to the observed CMBR dipole and quadrupole anisotropy are much larger than the Hubble radius at last scattering. Therefore, we will focus our analysis only on the evolution of the perturbation modes *outside the Hubble radius*.

We also wish to answer to another objection which could have been made against the model of PP, namely the fact that they consider our universe only in the phase of matter domination and that they add an ad hoc isocurvature perturbation at the time of the last scattering. A question of interest is whether some primordial isocurvature perturbations can survive during the evolution of the universe and be sufficiently important at the time of last scattering to produce effects comparable to those in PP. We show that this indeed the case by considering the influence of a pure isocurvature primordial perturbation on the dipole and quadrupole moments.

The plan of this paper is the following. In the section 2, we introduce the concept of adiabatic and isocurvature linear perturbations in a flat FRW background and we rederive the equations governing their evolution. In section 3, we give the expression for the anisotropy of the CMBR. In section 4, we make the connection between the Tolman-Bondi model used by PP and our cosmological perturbations approach. Finally we summarize in section 5 the observational implications of these results.

## II. ADIABATIC AND ISOCURVATURE PERTURBATIONS

There are several formalisms for dealing with cosmological linear perturbations. The oldest is due to Lifschitz [6] and uses the synchronous gauge. Another is the so-called gauge-invariant formalism of Bardeen [7], which employs arbitrary gauge and constructs gauge invariant quantities out of linear combinations of the perturbations. Finally there is also a covariant approach of cosmological perturbations, pioneered by Hawking [8] and developed recently by several authors [9] (see also references in [10]). We use here this latter formalism which we find more convenient: it employs quantities with a clear physical meaning, and in particular it provides a direct definition of the peculiar velocity, which turns out to be useful in interpreting the Sachs-Wolfe effect (see [11]).

We begin by reviewing this cosmological perturbation theory for a single fluid. Consider a space-time, endowed with a metric $g_{\mu\nu}$, filled with a perfect fluid with a number density, n, an energy density, $\rho$, a pressure, $p$ and a four-velocity $u^\mu$. One defines the comoving gauge as a particular foliation of space-time into hypersurfaces that are orthogonal to the matter flow $u^\rho$. This foliation is identified with the preferred foliation of the FRW spacetime (which we take to be flat for simplicity) representing the homogeneous approximation of the real spacetime. Note that such a foliation always exists in the linear approximation (whereas, in general, it requires a vorticity free flow).

We define a local Hubble parameter by

$$3H = \nabla_\rho u^\rho. \tag{1}$$

Using this definition, the local conservation of matter,

$$\nabla_\mu (n u^\mu) = 0, \tag{2}$$

can be written as

$$\frac{d\rho}{d\tau} = -3H(\rho + p), \tag{3}$$

where $d/d\tau$ is the derivative along the flow lines $u^\mu \nabla_\mu$. One can, then, write the Raychaudhuri equation, ignoring the terms involving the shear and the vorticity which are second order in the perturbations, as

$$\frac{dH}{d\tau} = -H^2 - \frac{4\pi G}{3}(\rho + 3P) - \frac{1}{3}\frac{D^2 p}{\rho + p}. \tag{4}$$

The operator, $D^2$, stands for $D_\mu D^\mu$ where $D_\mu$ is the covariant derivative projected on the hypersurface orthogonal to the flow $u^\mu$, i.e. $D_\mu = (g^\nu_\mu + u_\mu u^\nu)\nabla_\nu$. Equations 3 and 4 can be combined to a single equation governing the time evolution of $\delta = \delta\rho/\rho$:

$$\frac{d^2}{dt^2}\delta + H\left(2 + 3c_s^2 - 6w\right)\frac{d}{dt}\delta - \frac{3}{2}H^2\left(1 + 8w - 3w^2 - 6c_s^2\right)\delta = \frac{D^2 \delta p}{\rho}, \tag{5}$$

where $t$ is the time parameter of the comoving hypersurfaces. It is related to the proper time by ( [12])

$$\frac{d\tau}{dt} = \left(1 - \frac{\delta p}{\rho + p}\right). \tag{6}$$

The previous analysis deals only with a single perfect fluid. One can extend this treatment to several uncoupled perfect fluids (see [13] and [12]), each specified by its four-velocity $u_a^\rho$, its energy density $\rho_a$ and its pressure $P_a$. We are interested only in first order deviations from the FRW configuration where all the fluids have a common four-velocity. For each fluid one can define a Hubble parameter $H_a$ by an equation similar to (1). Moreover each fluid satisfied a conservation equation similar to (3) since the fluids are decoupled. It can then be shown [12] that the Raychaudhuri equation for each fluid becomes, ignoring higher order perturbative terms,

$$\frac{dH_a}{d\tau} + H_a^2 + \frac{4\pi G}{3}(\rho + 3p) = -\frac{1}{3}\frac{D^2 p_a}{\rho_a + p_a} + \frac{H - H_a}{\rho_a + p_a}\dot{p}_a. \tag{7}$$

By linearizing the conservation equation and the Raychaudhuri equation, one finds a system of coupled first order differential equations,

$$\frac{d}{dt}\delta_a = 3H_a\theta_a\frac{\delta p}{\rho_a} - 3(1 + w_a)\delta H_a + 3H_a\left(w_a - c_a^2\right)\delta_a, \tag{8}$$

and

$$\frac{d}{dt}\delta H_a = -2H\delta H_a - \frac{4\pi G}{3}\delta\rho - \frac{1}{3}\frac{D^2 p_a}{\rho_a + p_a} + \frac{\sum_a \theta_a \delta H_a - \delta H_a}{\rho_a + p_a}\dot{p}_a, \tag{9}$$

where one has defined

$$\theta_a = \frac{\rho_a + p_a}{\rho + p}, \qquad w_a = \frac{p_a}{\rho_a}. \tag{10}$$

A dot denotes the time derivation for the homogeneous background quantities.

If one considers only two fluids, it is useful to introduce the perturbation in the ratio of the number density

$$S = \frac{\delta_1}{1 + w_1} - \frac{\delta_2}{1 + w_2}, \tag{11}$$

and the total density perturbation

$$\delta = \frac{\delta\rho_1 + \delta\rho_2}{\rho_1 + \rho_2} = \frac{\rho_1}{\rho}\delta_1 + \frac{\rho_2}{\rho}\delta_2. \tag{12}$$

A general perturbation can be described by the pair $(\delta_1, \delta_2)$ or alternatively by the pair $(S, \delta)$. An adiabatic perturbation satisfies $S = 0$ and an isocurvature perturbation satisfies $\delta = 0$. Unfortunately, these conditions are not invariant with time and a perturbation that begins as an isocurvature perturbation generates an adiabatic component and vice versa. This was the origin of the claim of [5] that the analysis of PP is wrong. However, as we show later, and [5] fail to realize, for perturbations larger than the horizon, if $S \simeq 0$ initially $S$ will remain very small with respect to $\delta$ and vice versa. In particular this decomposition is meaningful for primordial fluctuations.

We now rewrite equations (8) and (9), adapted to the variables $(\delta_1, \delta_2)$, as evolution equations for the quantities $\delta$ and $S$. Using (8) and $\dot{w}_a = -3H(c_a^2 - w_a)(1 + w_a)$ (where $c_a^2 = \dot{P}_a/\dot{\rho}_a$), one finds that

$$\frac{d}{dt}S = -3(\delta H_1 - \delta H_2). \tag{13}$$

Differentiating this equation and using (9), we obtain a second order differential equation for $S$:

$$\frac{d^2 S}{dt^2} + \left(2 - 3c_z^2\right) H \frac{dS}{dt} = c_z^2 D^2 S + \left(c_1^2 - c_2^2\right) \frac{D^2 \delta}{1 + w}, \tag{14}$$

where $c_z^2 = c_1^2 \theta_2 + c_2^2 \theta_1$. Finally the equation for the evolution of the density perturbations $\delta$ follows from (5) when one expresses $\delta P$ in terms of $\delta$ and $S$:

$$\frac{d^2 \delta}{dt^2} + H\left(2 + 3c_s^2 - 6w\right)\frac{d\delta}{dt} - \frac{3}{2}H^2\left(1 + 8w - 3w^2 - 6c_s^2\right)\delta = c_s^2 D^2 \delta + (c_1^2 - c_2^2)\theta_1\theta_2(1 + w)D^2 S. \tag{15}$$

The physical situation, which we consider from now on, is the case where fluid 1 is pressureless ($c_1^2 = 0$) and fluid 2 is radiation ($c_2^2 = 1/3$). We define $a_{eq}$ as the scale factor at the matter-radiation transition, i.e. when the energy densities of the two fluids are equal. Then, by using $\rho_1/\rho_2 = a/a_{eq}$, one has

$$c_s^2 = \frac{1}{3}\left(1 + \frac{3}{4}\frac{a}{a_{eq}}\right)^{-1}, \qquad c_z^2 = \frac{1}{3}\left(1 + \frac{4}{3}\frac{a_{eq}}{a}\right)^{-1}, \qquad (16)$$

and

$$w = \frac{1}{3}\left(1 + \frac{a}{a_{eq}}\right)^{-1}, \qquad (c_1^2 - c_2^2)\theta_1\theta_2(1+w) = -\frac{1}{3}\left(1 + \frac{3}{4}\frac{a}{a_{eq}}\right)^{-1}\left(1 + \frac{a_{eq}}{a}\right)^{-1}. \qquad (17)$$

Note that with this choice of fluids 1 and 2 our definition of $S$ corresponds to the opposite of the variation of the specific entropy, i.e. the ratio of the number density of photons to the number density of dust (denoted $S$ in PP).

It is convenient to introduce the Fourier decomposition of the perturbations according to the definition

$$S_{\vec{k}} = \int \frac{d^3x}{(2\pi)^{3/2}} e^{-i\vec{k}.\vec{x}} S(\vec{x}), \qquad (18)$$

with a similar definition for $\delta_{\vec{k}}$. The evolution equations for the Fourier modes are simply equations (14) and (15) modified with the substitution of $-k^2/a^2$ in place of $D^2$. Each Fourier mode evolves independently of all the other modes, and one can thus study each mode individually. In standard cosmology the matter satisfies the strong energy condition $3p + \rho > 0$ and the comoving Hubble radius $(aH)^{-1}$ increases with time as the universe expands (the inverse is true during inflation). This implies that any given Fourier mode was outside the Hubble radius at sufficiently early time. Consequently, it is traditional, in standard cosmology, to define the initial conditions for the perturbations during the radiation dominated era at a stage when the relevant mode was outside the Hubble radius, i.e. when $k \ll aH$. During this stage, the r.h.s. of equation (14) can be neglected and the corresponding solutions are $S_{\vec{k}} \sim const.$ and $S_{\vec{k}} \sim \ln(t)$. The second solution is singular in the past and can be ignored. We see that primordial isocurvature perturbations are constant in time when they are outside the horizon:

$$S_{\vec{k}}(t) \simeq S_{\vec{k}}^{p}, \tag{19}$$

where the superscript 'p' denotes the primordial value. This remains true *even* after the transition between radiation domination and matter domination, as long as the modes are outside the Hubble radius (see Figure 1).

The isocurvature mode generates an adiabatic perturbation. We have integrated numerically the evolution of the density mode $\delta_k$ produced by a pure isocurvature primordial perturbation ($\delta_k = 0$). We find that $\delta_k$ grows. However, as long as the mode remains outside the Hubble radius, the value of $\delta_k$ is small with respect to the corresponding value of $S_k$. During the radiation era (see [10]),

$$\delta_k \sim \left(\frac{a}{a_{eq}}\right)^2 \left(\frac{k}{aH}\right)^2 S_k^p. \tag{20}$$

The amplitude of the density perturbation mode continues to grow after the radiation-matter transition, but as long as the mode is outside the Hubble radius, the amplitude is bounded from above by the asymptotic limit (see Fig. 2 and also [10])

$$\delta_k = \frac{2}{15}\left(\frac{k}{aH}\right)^2 S_k^p. \tag{21}$$

These results contradict the claim in [5] that the isocurvature perturbation is converted into an adiabatic perturbation after the equivalence. Our results are in agreement with other works (see [10] and references therein). To summarize, for the primordial isocurvature modes that remain outside the Hubble radius, $S_k$ remains constant during the whole evolution, whereas the density perturbations $\delta_k$, initially zero, grows like $a^4$ during the radiation era, then like $a$ during the matter era, while remaining small with respect to $S_k$. Once the perturbations enter the Hubble radius both $S_k$ and $\delta_k$ grow rapidly (see Fig. 1).

The same analysis for a primordial adiabatic perturbation reveals that, in the initial era, $\delta_k$ grows like $a^2$ while $S_k$, initially zero, evolves like

$$S_k \sim \frac{a}{a_{eq}}\left(\frac{k}{aH}\right)^2 \delta_k. \tag{22}$$

During the matter era, $\delta_k$ grows like $a$. We refer the reader to [10] for a more detailed discussion on adiabatic perturbations.

## III. THE COSMIC MICROWAVE BACKGROUND RADIATION

We turn now to the relation between the perturbations, discussed in section 2, and the observed CMBR anisotropy. The observed CMBR is the image of the last scattering surface which arrives today to our eyes (or in fact to our radio antennas). The fluctuations in the observed temperature are the consequence of both the perturbations in the matter content of the universe at the time of the last scattering and the perturbations of the geometry in the regions between the last scattering and us.

The temperature fluctuations due to the matter perturbations at the time of last scattering are intrinsic fluctuations. Since the radiation energy density is proportional to $T^4$ one always has:

$$\frac{\Delta T}{T} = \frac{1}{4}\frac{\delta \rho_r}{\rho_r}. \tag{23}$$

For an adiabatic perturbation ($S = 0$), (11) implies

$$\delta_r = \frac{4}{3}\delta_m. \tag{24}$$

The last scattering occurs during the matter dominated era, i.e. when $\rho_r < \rho_m$, hence $\delta \rho \simeq \rho \delta_m$ and therefore

$$\left(\frac{\Delta T}{T}\right)_{int} \simeq \frac{1}{3}\delta, \quad \text{adiabatic perturbation, matter era.} \tag{25}$$

For an isocurvature perturbation, $\delta \rho_r = -\delta \rho_m$. Hence during the matter era, $\delta_m$ can be neglected with respect to $\delta_r$ and (11) yields $S \simeq -(3/4)\delta_r$. Therefore

$$\left(\frac{\Delta T}{T}\right)_{int} \simeq -\frac{1}{3}S, \quad \text{isocurvature perturbation, matter era.} \tag{26}$$

The second source of the CMBR temperature fluctuations is the so-called Sachs-Wolfe effect [14]. It corresponds to the influence of the metric perturbations on the light rays during their travel between the last scattering surface and "our eyes". To express the Sachs-Wolfe effect we introduce the gravitational field $\psi$ and the peculiar velocity $\vec{v}$. The gravitational field, $\psi$, is given by the relativistic generalization of the Newtonian Poisson equation with a FRW background,

$$\frac{3}{2} H^2 \delta = D^2 \psi. \tag{27}$$

During the matter dominated era, as stated in section 2, the dominant solution for the density perturbation grows like $\delta \sim a$ and therefore the gravitational field is constant in time. Ignoring the decaying mode, the peculiar velocity field is defined as (see e.g. [10])

$$\vec{v} = -t \vec{D} \psi. \tag{28}$$

Using $\psi$ and $\vec{v}$ one can write the Sachs-Wolfe contribution to the CMBR anisotropy as

$$\left(\frac{\Delta T}{T}\right)_{SW} (\vec{e}) = \frac{1}{3} \left[\psi_{em} - \psi_0\right] + \vec{e}.\left[\vec{v}_{em} - \vec{v}_0\right], \tag{29}$$

where $\vec{e}$ is the unit vector corresponding to the direction of observation on the celestial sphere. The subscript $em$ means that the corresponding quantity is evaluated at the point on the last scattering surface that is observed today in the direction $\vec{e}$. The subscript 0 refers to the observer today.

The combination of equations (25) and/or (26) with equation (29) gives us the total CMBR anisotropy observed today (at least for large angular scales) for any configuration of the adiabatic and isocurvature perturbations, which is here specified in terms of the functions $\psi(r)$ and $S(r)$ at the time of last scattering.

The observed CMBR temperature fluctuations are generally decomposed in spherical harmonics :

$$\frac{\Delta T}{T}(\theta, \phi) = \sum_{l=1}^{\infty} \sum_{m=-l}^{l} a_{lm} Y_{lm}(\theta, \phi), \tag{30}$$

where $\theta$ (varying between 0 and $\pi$) and $\phi$ (varying between 0 and $2\pi$) are the usual angle coordinates on the two-sphere. One can express the individual coefficients of the decomposition as

$$a_{lm} = \int d\phi \, \sin\theta \, d\theta \, \frac{\Delta T}{T}(\theta, \phi) Y_{lm}^*(\theta, \phi), \tag{31}$$

In the particular spherically symmetric case, which we examine in the following, the temperature fluctuations depend only on the angle $\theta$, and all the coefficients with $m \neq 0$ vanish ( [15]). The non vanishing dipole and quadrupole coefficients are

$$a_{10} = \int_{-1}^{1} \frac{\Delta T}{T}(u) Y_{10}(u) du, \tag{32}$$

and

$$a_{20} = \int_{-1}^{1} \frac{\Delta T}{T}(u) Y_{20}(u) du, \tag{33}$$

with

$$Y_{10}(u) = \sqrt{\frac{3}{4\pi}} u, \qquad Y_{20}(u) = \sqrt{\frac{5}{4\pi}} \left(\frac{3}{2}u^2 - \frac{1}{2}\right), \tag{34}$$

where $u = \cos\theta$.

In the expression (31) the temperature anisotropy $\frac{\Delta T}{T}$ can also be seen as a function of $\vec{x}$ and can thus be decomposed in terms of its Fourier modes defined according to (18). We obtain

$$a_{lm} = \int \frac{d^3k}{(2\pi)^{3/2}} d\Omega_x Y_{lm}^*(\Omega_x) e^{i\vec{k}.\vec{x}} \left(\frac{\Delta T}{T}\right)_{\vec{k}}. \tag{35}$$

where $\Omega_x$ is the solid angle corresponding to $\vec{x}$. $x$ is the norm of $\vec{x}$ and represents the comoving distance between the observer and the last scattering surface. To a very good approximation, $x \simeq 2(H_0 a_0)^{-1}$ (the exact expression is given in (56)). Using then the identities

$$e^{i\vec{k}.\vec{x}} = 4\pi \sum_{l,m} i^l j_l(kx) Y_{lm}(\Omega_x) Y_{lm}^*(\Omega_k) \tag{36}$$

and

$$\int d\Omega \, Y_{l_1 m_1}(\Omega) Y^*_{l_2 m_2}(\Omega) = \delta_{l_1 l_2} \delta_{m_1 m_2}, \tag{37}$$

one finds

$$a_{lm} = 4\pi \int \frac{d^3 k}{(2\pi)^{3/2}} i^l j_l(kx) Y^*_{lm}(\Omega_k) \left(\frac{\Delta T}{T}\right)_{\vec{k}}. \tag{38}$$

For $kx \gg l(l+1)/2$,

$$j_l(kx) \sim \frac{1}{kx} \sin(kx - l\pi/2). \tag{39}$$

In the expression (38) the small scales are suppressed due to the presence of the factor $j_l(kx)$. A rough estimate of the coarse graining scale is $x$ for the dipole and $x/3$ for the quadrupole. In any case the modes that contribute to (32) and (33) are large scale modes, whose wavelength is far larger than the Hubble radius at the time of last scattering. The latter corresponds to an angular scale of $1°$ today. In practice we are thus allowed to work with the perturbations smoothed on a scale of the order of a few times the (comoving) Hubble radius at the last scattering. On one hand, these smoothed perturbations will not change the results in (32) and (33) at the notable exception of the dipole term due to $-\vec{e}.\vec{v}_0$. On the other hand, these smoothed perturbations contain only Fourier modes that remain outside the Hubble radius until the last scattering and are thus far easier to handle, as shown in the previous section.

To conclude this section we recall here that the satellite COBE [16] has measured the dipole and quadrupole components of the CMBR anisotropy to be:

$$D \equiv \sqrt{\frac{1}{4\pi} \sum_{m=-1}^{1} |a_{1m}|^2} \simeq 2 \times 10^{-3}, \tag{40}$$

$$Q \equiv \sqrt{\frac{1}{4\pi} \sum_{m=-2}^{2} |a_{2m}|^2} \simeq 5 \times 10^{-6}. \tag{41}$$

In fact the directly measured dipole is of the order $D \simeq 1.23 \times 10^{-3}$ but we have extrapolated this result, as is usually done, to get a dipole measured with respect to the Local Group frame [17].

## IV. THE TOLMAN-BONDI MODEL

The purpose of this section is to recover the numerical results of PP by an analytical method which is based on the linear theory presented in the previous sections. Indeed, although PP treated the complete non linear Tolman-Bondi (TB) problem, one can see from the order of magnitude of their quantities that one can study the same problem within the linear approximation. We begin, therefore, with the complete TB solution and we linearize it around the background solution of a flat FRW model.

The TB solution is given by the metric

$$ds^2 = -dt^2 + X^2(r,t)dr^2 + R^2(r,t)d\Omega^2, \tag{42}$$

where the coordinate $r$ is a comoving coordinate attached to an element of the dust fluid. The solution is characterized by two arbitrary functions of $r$. The first, $W(r)$, expresses how bound is a given shell $r$ with the binding energy being $W(r) - 1$. The second $t_s(r)$ describes the time in which the radius of a given shell vanishes and, following PP, we will consider here only solutions with $t_s = 0$. We define the gravitational mass as

$$m_b(r) \equiv \int_0^r 3r^2 W(r) dr. \tag{43}$$

The metric functions $X$ and $R$ satisfy then

$$X = \frac{R'}{W(r)}, \tag{44}$$

$$\dot{R}^2 = W^2(r) - 1 + \frac{2m_b(r)}{R}, \tag{45}$$

where a dot denotes a partial derivative with respect to the time coordinate $t$ and a prime denotes a partial derivative with respect to the radial coordinate $r$. The mass density $\rho_m(r,t)$ is given by the expression

$$\rho(r,t) = \frac{3r^2}{4\pi X R^2}. \tag{46}$$

Our background model is a dust dominated flat FRW universe, corresponding to the particular TB solution with $W(r) = 1$. One can easily solve (44) and (45) in this case, and with the initial condition $R(r, 0) = 0$, one finds

$$X = a(t), \qquad R = a(t)r. \tag{47}$$

The FRW scale factor $a(t)$ is explicitly given by

$$a(t) = \alpha t^{2/3}, \qquad \alpha = (9/2)^{1/3}. \tag{48}$$

Consider now a small deviation from this flat FRW model:

$$W(r) = 1 + \delta W(r). \tag{49}$$

Then the linearization of equations (44) and (45) yields

$$\delta X = \delta R' - a \delta W, \tag{50}$$

and

$$\dot{a} r \delta \dot{R} + \frac{r}{a^2} \delta R = \delta W + \frac{2}{ar} \int 3r^2 \delta W \, dr. \tag{51}$$

The second term on the right hand of this last equation decays like $a^{-1}$ with respect to the first one. Dropping it, one can easily find the dominant solution:

$$\delta R = \frac{9}{10\alpha} \frac{\delta W}{r} t^{4/3}. \tag{52}$$

The dominant solution for $\delta X$ follows immediately:

$$\delta X = \frac{9}{10\alpha} \left( \frac{\delta W}{r} \right)' t^{4/3}. \tag{53}$$

One can also obtain the dominant solution for the matter energy density $\delta_m$:

$$\delta_m = -\frac{9}{10\alpha^2} \frac{(r\delta W)'}{r^2} t^{2/3} = -\frac{9 t_0^{2/3}}{10\alpha^2} (1+z)^{-1} \frac{(r\delta W)'}{r^2}. \tag{54}$$

Substitution of expression (54) for $\delta_m$ in the Poisson equation (27) and using (48) yields an expression for the gravitational potential:

$$\psi(r) = -\frac{3}{5r} \int dr \frac{(r\delta W)'}{r}. \tag{55}$$

$\psi$ is time independent as expected.

It remains to calculate the temperature anisotropy for this spherical symmetric model. Consider an observer located at $r_0$ at time $t_0$. There is a preferred axis which links this observer to the center $r = 0$. Since there is an axial symmetry around this axis, one can restrict oneself to the meridional plan. One then denotes $\theta$ the angle between the preferred axis and the direction of observation ($\theta$ varies between 0 and $\pi$). Let $d$ be the comoving distance between the observer and the last scattering surface:

$$d = 2H_0^{-1} a_0^{-1} \left[1 - (1+z)^{-1/2}\right] = 3\alpha^{-1} t_0^{1/3} \left[1 - (1+z)^{-1/2}\right], \tag{56}$$

where $z$ is the redshift corresponding to the last scattering surface (for the numerical application, we shall take $1 + z = 1000$). Then the radial distance of a point of the last scattering surface corresponding to the angle of observation $\theta$ is given by

$$r^2 = d^2 \sin^2\theta + (r_0 + d\cos\theta)^2. \tag{57}$$

Finally, using (29) and (28) the temperature anisotropy of the CMBR is given by

$$\frac{\Delta T}{T}(\theta) = \left(\frac{\Delta T}{T}\right)_{int}(r_{em}) + \frac{1}{3}[\psi(r_{em}) - \psi(r_0)] + \left(\frac{t}{a}\right)_{em} \frac{d\psi}{dr}(r_{em}) \frac{r_0\cos\theta + d}{r} - \left(\frac{t}{a}\right)_0 \frac{d\psi}{dr}(r_0)\cos\theta. \tag{58}$$

If the function $\Delta T/T$ varies sufficiently slowly one can expand it in a Taylor series and evaluate analytically the integrals (32) and (33). Starting with the Sachs-Wolfe part of the temperature fluctuation, one can see that there is a cancellation in the non symmetric terms (i.e. the dipole terms) at the order $(d^2/r_0^2)$ and that one must therefore go to the following order $(d^3/r_0^3)$, which also implies that terms proportional to $u^3$ will appear. One finally obtains

$$\left(\frac{\Delta T}{T}\right)_{SW}(u) \simeq D_{1SW}u + D_{2SW}u^3 + Q_{SW}u^2, \tag{59}$$

with

$$D_{1SW} = \left[-\frac{1}{2}r_0\frac{d\psi}{dr}(r_0) + \frac{r_0^2}{2}\frac{d^2\psi}{dr^2}(r_0)\right]\frac{d^2}{r_0^2}\left(\frac{d}{3} + 3\alpha^{-1}t_0^{1/3}(1+z)^{-1/2}\right) + \mathcal{O}\left(\frac{d^4}{r_0^4}\right), \tag{60}$$

$$D_{2SW} = \left[\frac{1}{2}r_0\frac{d\psi}{dr}(r_0) - \frac{r_0^2}{2}\frac{d^2\psi}{dr^2}(r_0) + \frac{r_0^3}{6}\frac{d^3\psi}{dr^3}\right]\frac{d^2}{r_0^2}\left(\frac{d}{3} + 3\alpha^{-1}t_0^{1/3}(1+z)^{-1/2}\right) + \mathcal{O}\left(\frac{d^4}{r_0^4}\right) \tag{61}$$

and

$$Q_{SW} = d\left[\frac{d^2\psi}{dr^2}(r_0) - r_0^{-1}\frac{d\psi}{dr}(r_0)\right]\left(\frac{d}{6} + \alpha^{-1}t_0^{1/3}(1+z)^{-1/2}\right) + \mathcal{O}\left(\frac{d^3}{r_0^3}\right). \tag{62}$$

The intrinsic temperature fluctuation is simpler to evaluate. If we note $(\Delta T/T)_{int} = I$, where $I$ is either $\delta/3$ in the adiabatic case and $-S/3$ in the isocurvature case, then the Taylor expansion yields

$$\left(\frac{\Delta T}{T}\right)_{int}(u) \simeq D_I u + Q_I u^2, \tag{63}$$

with

$$D_I = \frac{dI}{dr}(r_0)d + \mathcal{O}\left(\frac{d^2}{r_0^2}\right), \tag{64}$$

$$Q_I = \frac{d^2}{2}\left[\frac{d^2I}{dr^2}(r_0) - r_0^{-1}\frac{dI}{dr}(r_0)\right] + \mathcal{O}\left(\frac{d^3}{r_0^3}\right). \tag{65}$$

Finally the total dipole and quadrupole, expresses in terms of the above coefficients, are

$$D = \frac{1}{\sqrt{3}}\left(D_I + D_{1SW} + \frac{3}{5}D_{2SW}\right), \tag{66}$$

and

$$Q = \frac{2}{3\sqrt{5}} \left( Q_I + Q_{SW} \right). \tag{67}$$

The expression seems similar for both the isocurvature and adiabatic perturbations. However, the Poisson equation (27) with the expression (56) shows that $\delta$ is always of the order $(d/r_0)^2 (1+z)^{-1}$ times $\psi$. Therefore the intrinsic dipole contribution due to $\delta$ is smaller than the Sachs-Wolfe dipole by a factor of the order $(1+z)^{-1}$ and is thus always negligible. But when there is an isocurvature perturbation, one sees, in contrast, that the intrinsic dipole will be in general the dominant term.

### A. PP results

We are now in position to compute analytically the results of PP. Their specific model corresponds here to

$$\delta W(r) = -\frac{1}{2} \frac{1-r^2}{1+r^2} r^2, \tag{68}$$

and

$$S(r) = \frac{1}{2} \frac{r^2}{1+r^2}. \tag{69}$$

It turns out that this expression for $\delta W(r)$ yields an explicit expression for $\psi(r)$:

$$\psi(r) = \frac{3}{10} \left[ \frac{1-r^2}{1+r^2} r - \frac{r}{2} + \frac{\ln(1+r^2)}{r} \right]. \tag{70}$$

The numerical results are the following (with $t_0 = 10^{-6}$):

$$D_I(iso) \simeq -1.47 \times 10^{-3}, \quad D_{1SW} \simeq -1.26 \times 10^{-7}, \quad D_{2SW} \simeq 5.09 \times 10^{-7} \tag{71}$$

$$Q_I(iso) \simeq 2.58 \times 10^{-5}, \quad Q_{SW} \simeq -6.94 \times 10^{-6}. \tag{72}$$

For the model without isocurvature perturbation, one finds $D \simeq 1.0 \times 10^{-7}$ and $Q \simeq 2.1 \times 10^{-6}$, whereas for the model with isocurvature perturbation, one finds $D \simeq 8.5 \times 10^{-4}$ and $Q \simeq 1.9 \times 10^{-5}$. These numerical results should be compared with the results obtained by numerical integration of the light rays, and given in the figures 3 and 4 of PP. Our results here are limited to $r_0 = 1$ since we have assumed a flat space from the beginning. The numerical values given above correspond to $t_0 = 10^{-6}$ but the corresponding values for a different $t_0$ can be obtained immediately by noticing that the dependence on $t_0$ of $D_I$ is due to an overall multiplicative term $t_0^{1/3}$ whereas $D_{1SW}$ and $D_{2SW}$ are proportional to $t_0$, and $Q_{SW}$ and $Q_I$ proportional to $t_0^{2/3}$. Comparison with PP shows a good agreement, thus confirming the conclusions of PP, although there are small discrepancies between the precise numerical values, which we cannot explain.

One could have argued against the results of PP that their two perturbations are a priori completely independent. The question arises what will happen if one considers an initial set of primordial perturbations and let it evolve in time until the time of last scattering. Will it be possible to reproduce similar results with these more stringent conditions? In [5], it is argued that it will be impossible to recover PP results. What we show in the next subsection is that indeed we can recover the same behavior.

### B. Primordial isocurvature perturbation

We turn now to consider the extreme case where *primordial* isocurvature perturbations on extremely large scale (i.e. scales much larger than the horizon) are the only source of the observed dipole and quadrupole. We denote by $S_k^p$ the modes of the primordial isocurvature perturbation ($k < a_0 H_0$). It follows from the analysis of Section 2 that:

$$S_k(t_{ls}) \simeq S_k^p \tag{73}$$

where $ls$ stands for the last scattering. The primordial isocurvature perturbation has also generated a energy density perturbation given by

$$\delta_k = \frac{2}{15} \left( \frac{k}{aH} \right)^2 S_k^p, \tag{74}$$

and a corresponding gravitational potential:

$$\psi_k = -\frac{1}{5}S_k^p. \tag{75}$$

We choose, as an example, a primordial isocurvature perturbation of the form

$$S^p(r) = a_s \frac{(r/r_s)^2}{1 + (r/r_s)^2}. \tag{76}$$

One can adjust the two parameters of the perturbation, namely the amplitude $a_s$ and the wavelength $r_s$, to reproduce exactly the measured dipole and quadrupole. One finds $a_s \simeq 2.94$ and $r_s \simeq 2.80$, which means that the wavelength of the perturbation must be roughly 150 times larger than the Hubble radius today. Note that such adjustment is possible only for isocurvature modes. This would be impossible with an adiabatic mode, as will be shown clearly in the next section.

## V. INTERPRETATION OF THE RESULTS

Before turning to observational implications, we wish to extract, in this section, the essential arguments which explain why the isocurvature perturbation and the adiabatic perturbation produce such different results. To show this we consider a plane wave perturbation and obtain rough estimates for all terms involved. We show in particular that the ratio $D/Q$ for a very large scale perturbation is inverted when one goes from an adiabatic to an isocurvature perturbation.

### A. Adiabatic perturbations

We have already emphasized that the main contribution to the dipole or even to the quadrupole arises from large scales. We shall restrict, therefore, our analysis to the Fourier modes outside the Hubble radius (at the time of the last scattering). The Fourier transform of the relativistic Poisson equation (27) reads

$$\delta_k = -\frac{2}{3}\left(\frac{k}{aH}\right)^2 \psi_k. \tag{77}$$

Assuming a monochromatic perturbation of the form $\delta \sim \delta_k e^{ikx}$, the intrinsic dipole contribution is, therefore, of the order

$$D_I\{\delta\} \sim \left(\frac{k}{a_0 H_0}\right)\delta_k \sim \left(\frac{k^2}{a^2 H^2}\right)_{ls} \left(\frac{k}{aH}\right)_0 \psi_k. \tag{78}$$

The dipole contribution of $\psi/3$ is given by ( [11])

$$D\{\frac{1}{3}\psi\} = - <\vec{v}(t_0) - \vec{v}(t_{ls})>, \tag{79}$$

where the average is taken over the comoving volume defined as the intersection of our past light-cone with the last scattering hypersurface

$$<\vec{v}(t)> = V_{ls}^{-1} \int_{ls} d^3x\, \vec{v}(t,\vec{x}). \tag{80}$$

Finally, we recall that the peculiar velocity evolves like $t/a$. Thus, the peculiar velocity at the last scattering is negligible with respect to the peculiar velocity today and the main contribution to the Sachs-Wolfe dipole for adiabatic perturbations is

$$D_{SW} \simeq \vec{e}.\left(\vec{v}_0 - <\vec{v}>(t_0)\right). \tag{81}$$

This result corresponds to the standard statement that the dipole is due to the relative motion of the Earth. Note that this calculation gives a precise definition of the relative velocity of the Earth and in particular with respect to which frame.

Using a Taylor expansion within the integral in (80) one obtains:

$$D_{SW} \simeq \vec{e}.\left(\vec{v}_0 - <\vec{v}>(t_0)\right) \sim \left(\frac{k^2}{a_0^2 H_0^2}\right) v_k(t_0) \sim \left(\frac{k}{a_0 H_0}\right)^3 \psi_k. \tag{82}$$

$\vec{v}_0$ corresponds, in this formula, to our peculiar velocity induced only by the single very large scale mode under consideration. If one takes into account the contribution of small scale modes to our peculiar velocity then one sees that the net measured velocity (and hence the measured dipole) will be dominated by the contribution of the sub-horizon modes. Moreover, comparison of (78) with (81) and (82) shows that $D_I\{\delta\} \sim (1+z)^{-1}D_{SW}$ and therefore the total dipole is $D \simeq D_{SW}$.

The expected dipole can be evaluated given a power spectrum (the adiabatic perturbations are supposed to be distributed like a Gaussian random field). Using the power spectrum $\mathcal{P}_\psi$ of $\psi$ defined by

$$\langle \psi_{\vec{k}} \psi_{\vec{k}'}^* \rangle = 2\pi^2 k^{-3} \mathcal{P}_\psi(k) \delta(\vec{k}-\vec{k}'), \tag{83}$$

we find, taking into account only the (dominant) term $\vec{e}.\vec{v}_0$,

$$\langle |a_{10}|^2 \rangle = \frac{16\pi}{27} \int \frac{dk}{k} \left( \frac{k^2}{(a_0 H_0)^2} \right) \mathcal{P}_\psi(k). \tag{84}$$

For the most common spectrum, namely the scale invariant Harrison-Zeldovich spectrum, $\mathcal{P}_\psi$ is a constant and we express it as a function of the expected quadrupole

$$\mathcal{P}_\psi = \frac{27}{\pi} \Sigma_2^2, \tag{85}$$

with $\Sigma_2^2 = \langle |a_{2m}|^2 \rangle$. The contribution to the rms-dipole of the fluctuations of scales larger than a given scale is obtained by integrating the above expression (84) with a lower cut-off $\bar{k}$. Hence

$$\langle v^2 \rangle^{1/2} = \sqrt{\frac{3}{4\pi}} \langle |a_{10}|^2 \rangle^{1/2} = \sqrt{\frac{6}{\pi}} \Sigma_2 \frac{\bar{k}}{a_0 H_0}. \tag{86}$$

This gives for the peculiar velocity

$$v_{rms} \simeq 250 km\ s^{-1} \left( Q_{rms-PS}(n=1)/17\mu K \right) \left( \lambda/50 h^{-1} Mpc \right)^{-1}, \tag{87}$$

where $h$ is the Hubble constant in units 100 km s$^{-1}$Mpc$^{-1}$ and $\lambda$ is the smoothing scale for the fluctuations (the relation between $Q_{rms-PS}(n=1)$ such as it is defined in [16] and $\Sigma_2$ is $\Sigma_2 = \sqrt{4\pi/5}(Q_{rms-PS}/T_0)$, where $T_0 = 2.73K$ is the CMBR monopole temperature). Instead of using a cut-off in the Fourier space, one may wish to introduce a cut-off in the real space, i.e. use a top hat window function, in which case the expression (86) should be multiplied by the factor $\sqrt{4.5} \simeq 2$. This relation shows that our velocity measured with respect to a given frame is inversely proportional to the distance of this frame from us.

The CMBR quadrupole is given by the quadrupole part of $\psi/3$ which is of the order

$$Q_{ad} \sim \left(\frac{k^2}{a_0^2 H_0^2}\right) \psi_k. \tag{88}$$

The expected ratio between the dipole and quadrupole contributions arising from a single mode very large scale adiabatic perturbation is, therefore, of the order

$$\left(\frac{D}{Q}\right)_{ad} \sim \left(\frac{k}{a_0 H_0}\right). \tag{89}$$

This is less than unity, by definition, hence the observed dipole and quadrupole cannot be explained by such a perturbation. In the context of adiabatic perturbations, the origin of the dipole must be *only* peculiar velocity.

Note finally that, since $aH = (1+z)^{1/2} a_0 H_0$ and $aH \sim t_0^{-1/3}$ in the matter dominated era, expressions (82) and (88) show that $D \sim t_0$, $Q \sim t_0^{2/3}$ and $(D/Q)_{ad} \sim t_0^{1/3}$. This is in agreement with the *numerical* dependence observed by PP.

### B. Isocurvature perturbations

We turn now to isocurvature perturbations. Whereas the dependence on the gravitational potential $\psi$ of the Sachs-Wolfe term due to a primordial isocurvature perturbation is the same as that due to a primordial adiabatic perturbation, the intrinsic contribution is drastically different. In particular the dipole of the intrinsic anisotropy,

$$D\{-\frac{1}{3}S\} \sim \left(\frac{k}{a_0 H_0}\right) \psi_k, \tag{90}$$

is dominant with respect to the Sachs-Wolfe dipole for scales larger than the Hubble radius today. The total quadrupole, on the other hand, is comparable to the adiabatic one:

$$Q_{iso} \sim \left(\frac{k^2}{a_0^2 H_0^2}\right)\psi_k. \tag{91}$$

Combining the last two expressions, we find that the expected ratio between the dipole and quadrupole is

$$\left(\frac{D}{Q}\right)_{iso} \sim \left(\frac{k}{a_0 H_0}\right)^{-1}. \tag{92}$$

Once more, a comparison with the numerical behaviours observed in PP is instructive. The quadrupole $Q$ has the same form as in the adiabatic case: $Q \sim t_0^{2/3}$. The dipole is different and it follows from the above expression that it evolve like $D \sim t_0^{1/3}$. The ratio behaves now like $(D/Q)_{iso} \sim t_0^{-1/3}$. All these dependences are confirmed by the numerical results of PP.

### C. Continuous isocurvature spectrum

So far we have dealt only with ultra large scale monochromatic perturbations. One can wonder what will be modified when one considers a continuous spectrum instead of a single mode. We examine here only a pure isocurvature primordial spectrum. The case of an adiabatic power spectrum is extensively treated in the literature (see e.g. [10]).

We consider primordial isocurvature perturbations that are described by a homogeneous and isotropic Gaussian random field, which is completely specified by its power spectrum $\mathcal{P}_S(k)$:

$$\langle S_{\vec{k}} S_{\vec{k}'}^* \rangle = 2\pi^2 k^{-3} \mathcal{P}_S(k) \delta(\vec{k} - \vec{k}'). \tag{93}$$

In particular one can compute the variance of the distribution of the multipoles as a function of the power spectrum $\mathcal{P}_S$:

$$\Sigma_l^2 = \langle |a_{lm}|^2 \rangle = \frac{4\pi}{9} \int \frac{dk}{k} \mathcal{P}_S(k) j_l^2(2k/a_0 H_0), \tag{94}$$

where it is assumed that the intrinsic contribution is the dominant one in the temperature anisotropy. The dipole and quadrupole correspond roughly to $\Sigma_1$ and $\Sigma_2$ respectively. We now assume that the power spectrum is a power law $\mathcal{P}_S(k) \simeq A k^n$, which is a standard assumption made in cosmology, and moreover that this power spectrum as an upper cut-off $\bar{k}$.

If the cut-off $\bar{k}$ is such that $\bar{k} \ll a_0 H_0$, then it is legitimate to use the approximation

$$j_l(x) \simeq \frac{x^l}{(2l+1)!!}, \qquad x \ll 1. \tag{95}$$

One can then calculate explicitly the dipole and the quadrupole:

$$D \sim \bar{k}^{(n+1)/2} (a_0 H_0)^{-1}, \qquad Q \sim \bar{k}^{(n+3)/2} (a_0 H_0)^{-2}. \tag{96}$$

The ratio between the dipole and quadrupole due to a very large scale (power-law) power spectrum should thus be of the order

$$\frac{D}{Q} \sim \left( \frac{\bar{k}}{a_0 H_0} \right)^{-1}, \tag{97}$$

which is of the same order as in the case of a monochromatic perturbation.

If there is no cut-off in the power spectrum or if the cut-off is smaller than the Hubble radius today, i.e. $\bar{k} \geq a_0 H_0$, then the dominant contribution in the integral (94) both for $l=1$ and $l=2$ will come from the wavelengths roughly of the same order than the Hubble radius (today), as explained at the end of section 3. Therefore the corresponding dipole and quadrupole should be of the same order of magnitude in this case. Beware that the dipole and quadrupole here are computed by taking into account only the intrinsic contribution. For the scales smaller than the Hubble radius (today), the primordial isocurvature modes have produced adiabatic perturbations with corresponding gravitational potential and peculiar velocity field. The dipole can therefore be dominated by the Doppler effect due to our peculiar velocity, as in the standard interpretation of the dipole, and we thus find that the dipole can be large relative to all other multipoles, even in this case.

Finally we must mention the hybrid possibility that the observed dipole could result of a combination of a Doppler effect (due to either adiabatic or isocurvature primordial perturbations) and of an intrinsic ultra large scale isocurvature contribution.

## VI. CONCLUSION

We have shown that it is possible that the observed CMBR dipole, or a significant fraction of it, has a non Doppler origin. This can happen if there is a suitable ultra large scale (typically beyond 100 times the size of the present Hubble radius) isocurvature perturbation. That is an ultra large scale fluctuation in the ratio of photons to baryons. This fluctuation will induce predominantly a dipole component in the observed horizon, without inducing higher order multipoles. This fluctuation should not be accompanied by isocurvature perturbations on smaller scales (between $100 H_0^{-1}$ and $0.1 H_0^{-1}$), because these would induce higher order moments which would then be comparable to the dipole.

It is clear from the above analysis that if the observed CMBR has a non Doppler origin then there should be a unique mechanism that would produce these very large scale isocurvature perturbations and will distinguish them from the rest of the power spectrum. A priori there are several possible origins for these ultra large scale perturbations. They could have been produced during an inflation era: the model of inflation (see e.g. [18]) must then include several scalar fields in order to allow for isocurvature perturbations in addition to the always present adiabatic ones. Another explanation would be that these perturbations are the remnants of the preinflationary epoch of the universe, as was suggested by Turner [4]. Indeed, if the duration of inflation is slightly more than what is needed to solve the "horizon" problem, then the scales that were larger than the Hubble radius at the onset of inflation would be today also larger than the Hubble radius but not by a large amount.

Is it possible to distinguish between an isocurvature dipole and a dipole due to our peculiar velocity? Luckily, there is a clear direct observational test. We have seen in the last section that the dominant contribution to our peculiar velocity arises from small scale modes and hence it should converge to the same velocity when it is measured relative to different distant frames. Thus a peculiar velocity dipole will induce the same dipolar pattern in other background fields, such as the X-ray background or $\gamma$-ray bursts which are located at $z \approx 1 - 2$. Depending of the power spectrum even nearer frames such as optical galaxies, IRAS galaxies, distant supernovae and Abell clusters should display similar peculiar motion pattern. Note however, that for distances smaller than the horizon the intrinsic contribution to the dipole might not be negligible and fluctuations in the density of sources should be included [19]. A convergence of all those peculiar velocities will support the Doppler origin of the CMBR. A final confirmation should arise if the observed peculiar motion is consistent with the expected r.m.s. value of this quantity, given by Equation (86), as calculated from the observed power spectrum of the matter fluctuations (one has of course to be careful here about cosmic variance). If this interpretation is confirmed then the measured quadrupole shows that the Universe is homogeneous at least on scales that are larger by $10^5$ than the current horizon. This will immediately rule out the existence of significant ultra large scale isocurvature perturbations and cosmological scenarios that produce them.

If, on the other hand, the observed dipole is due to extremely large scale isocurvature fluctuations, it should not have any corresponding signature on small scales. We expect, in this case that the observed dipole relative to the nearer frames, mentioned earlier, will still converge, but now to a different velocity in both magnitude and direction than the velocity implied by the CMBR dipole. We should point out that failure of the peculiar velocity to converge on those nearer scale would imply that the primordial power spectrum has some peculiar behavior on intermediate scales, a behavior that causes the integral in equation (94) to fluctuate.

The observed CMBR dipole implies a velocity of the local group of $627 \pm 22$ km s$^{-1}$ and it points towards the galactic coordinates ($l = 276^o \pm 3^o, b = 33^o \pm 3.^o$). The magnitude of this velocity is larger than the expected r.m.s. value for a Harrison-Zel'dovich spectrum normalized by COBE, which is several hundred km s$^{-1}$. The measured dipole in the X-ray background [20], which arises from sources at $z \approx 1.5$ is within the statistical errors. The dipole has not been measured relative to any other sources at comparable distances.

However, it has been measured relative to nearby galaxies [21], IRAS galaxies [22] and distant supernovae [23] which are all at $z < 0.03$. The COBE observations are within the statistical errors of all those measurements. This suggests that we are observing the convergence of the dipole on smaller scales. The observed dipole in distant Abell clusters [24] whose magnitude is $561 \pm 284$ km s$^{-1}$ towards ($l = 220^o \pm 27^o, b = -28^o \pm 27^o$) is, however, inconsistent with the CMBR dipole. It is also inconsistent with the distant supernovae dipole [23] which is measured relative to objects at the same distances. Hence we can conclude that at present the observational situation tends towards the conventional Doppler origin but the situation is inconclusive yet. Further measurement in the future of dipole relative to additional frames or refinement of current measurements should provide a conclusive answer in the future.

## ACKNOWLEDGMENTS

D.L. was supported by a Golda Meir postdoctoral fellowship.

FIGURES

Fig. 1: *The amplitude of the entropy perturbation ($S_k$) and the density perturbation ($\delta_k$) as a function of $\log(a/a_{eq})$, for an initial pure isocurvature perturbation with an initial amplitude $10^{-2}$. Three different wavelengths are shown. The first (solid curve for S and dotted curve for $\delta$) remains always larger than the horizon. The second (short-dashed curve for S and long-dashed curve for $\delta$) enters the horizon at $a \approx 1.5 a_{eq}$, which is marked in the figure by a square. The third perturbation (dashed-dotted curve for S and curve dashed-dotted line for $\delta$) enters the horizon at $a \approx 0.05 a_{eq}$, which is marked by a triangle.*

Fig. 2: *The quantity $(aH/k)^2 \delta_k$ as a function of $\log(a/a_{eq})$ for a primordial pure isocurvature perturbation with five different wavelengths. The two long wavelengths (dotted curve and long-dashed-short-dashed curve) overlap. These modes do not enter the horizon and they approach the asymptotic limit (straight solid curve) 2/15. An intermediate wavelength perturbation (short dashed curve) deeps slightly before going up after it has entered the horizon at $a \approx a_{eq}$. The same behaviour is seen in the two short wavelength perturbations (solid curve and long-dashed curve) which deep first and then grow rapidly after entering the horizon at $a < 0.1 a_{eq}$.*

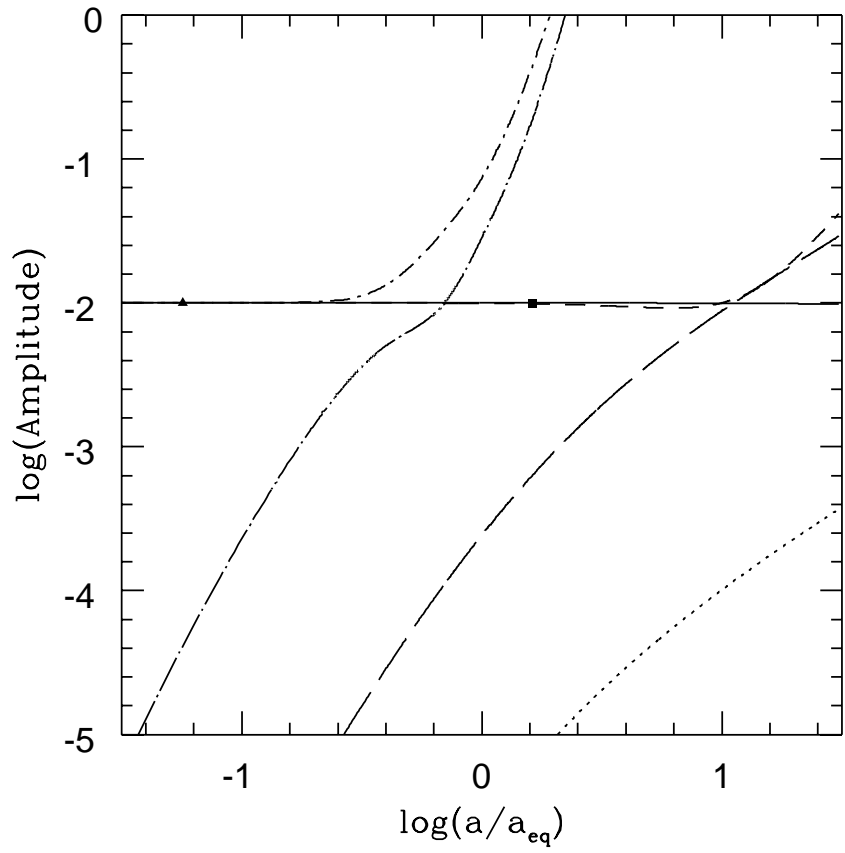

Fig. 1

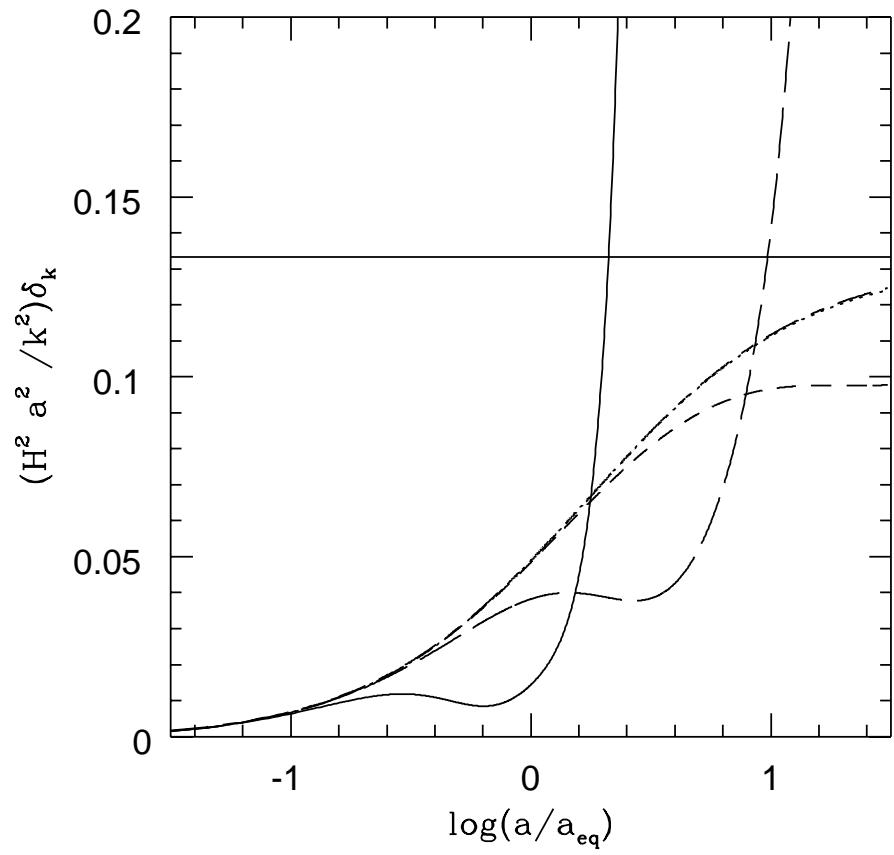

Fig. 2